# Strategic SDN-based Microgrid Formation for Managing Communication Failures in Distribution System Restoration

Jian Zhong, Chen Chen, *Senior Member, IEEE,* Zhaohong Bie, *Fellow, IEEE,* and Mohammad Shahidehpour, *Life Fellow, IEEE*

***Abstract*—Grid modernization has increased the reliance of power networks on cyber networks within distribution systems (DSs), heightening their vulnerability to disasters. Communication network failures significantly impede DS load recovery by diminishing observation and control. Prior research has largely ignored the need for integrated recovery of DS power and cyber networks' centralized control. Indeed, communication network restoration is critical for speedy load recovery through DS automation based microgrid formation. This paper exploits the data routing capabilities of software-defined networking (SDN) to enhance centralized control recovery in DS communication networks, incorporating it into a comprehensive DS restoration model. This model, tailored to the control requirements of load restoration, strategically allocates limited communication resources to re-establish connections between the operation center and terminal devices. Subsequently, DS automation is employed to orchestrate DS microgrid formation for power resupply. Additionally, we introduce a cyclic algorithm designed to optimize the load recovery via a multi-step, cooperative process. The efficacy of the proposed method is demonstrated on IEEE 33-node and IEEE 123-node test feeders.**

***Index Terms*—Distribution system restoration, cyber-physical resiliency, communication recovery, software-defined networking (SDN), microgrid formation.**

## NOMENCLATURE

*A. Indices and Sets*

| | |
|---|---|
| $i, j$ | Index for DS buses |
| $k$ | Index for DS lines |
| $l$ | Index for communication links |
| $m, n$ | Index for communication nodes |
| $\mathcal{L}^C$ | Set of communication links |
| $\mathcal{L}^P$ | Set of DS lines |
| $\mathcal{N}^C$ | Set of communication nodes |
| $\mathcal{N}^P$ | Set of DS buses |
| $\mathcal{S}^p$ | Set of buses with power sources; |

*B. Parameters*

| | |
|---|---|
| $\chi_i^{PN}, \chi_k^{PL}$ | Equipment states of bus $i$ and line $k$ |
| $\chi_m^{CN}, \chi_l^{CL}$ | Equipment states of node $m$ and link $l$ |
| $\xi_l^L$ | Propagation delay of link $l$ |
| $\xi_m^N$ | Forwarding delay of node $m$ |

*C. Variables*

| | |
|---|---|
| $b_k$ | Operational state of line $k$ |
| $d_m^N, d_l^L$ | Consumed bandwidth of node $m$ and link $l$ |
| $e_m$ | End-to-end data delay from terminal device $m$ to the operation center |
| $f_i^N$ | Electrified state of bus $i$ |
| $f_k^L$ | Commodity flow from bus $i$ to bus $j$ on line $k = (i,j)$ |
| $h_i^m, h_i^l$ | 0-1 variables indicating whether the data of terminal device $i$ flow through node $m$ and link $l$ |
| $s_m$ | 0-1 variable indicating the communication state of node $m$ to the operation center |

## I. INTRODUCTION

WITH global advancements in grid modernization, advanced information and communication technologies (ICTs) are being increasingly integrated into distribution systems (DSs). Consequently, DSs have transformed into cyber-physical systems where communication and power networks are tightly interdependent [1], as illustrated in Fig. 1. However, this interdependence makes DSs vulnerable to both physical and cyber failures [2]. A failure in one network can trigger failures in the other, potentially leading to cascading failures [3]. For instance, on September 28, 2003, Italy suffered a major blackout due to a cascade of cyber-physical failures, starting with power plant shutdowns and resulting in communications network node failures, which further disrupted power stations [4]. The 2011 Tohoku earthquake in Japan revealed how damage to cellular sites impeded utility repair crew communication, significantly hindering grid recovery [5]. Moreover, cyber-physical breakdowns during extreme events intensify failure progression, leading to more outages and amplified system losses [6]. Therefore, there is an increased demand for effective DS restoration strategies to address both cyber and physical failures.

The key to DS resilience involves leveraging various flexible resources to maintain electricity supply for consumers [2]. A crucial method for power restoration after outages is to reconfigure the network topology to dynamically form microgrids with distributed generators (DGs) [7], thus minimizing outage impacts. Indeed, manual on-site operations, including traveling to disaster sites of DSs, are compromised by natural hazards such as icy roads post-snowstorms [8] and flooded roads following heavy rains [9]. These conditions significantly slow down travel, delay restoration efforts, and potentially endanger staff safety [10]. DS automation rapidly facilitates this process by controlling automated switches. It provides speed, automation, and centralization in DS restoration, thus reducing manual operation risks during extreme events [11]. To optimize microgrid formation, techniques such as mixed-integer nonlinear programming (MINLP) [12], mixed-integer linear programming (MILP) [13], and heuristic search algorithms [2] have been used. These methods enable effective control over DSs to form multiple microgrids, aiming to restore the maximum possible load post-blackout. However, traditional approaches often do not consider or idealize the DS's communication infrastructure as intact [7, 13] which differs from the actual scenario. In fact, due to geographical distribution [14], DS communication networks in most cases are spatially bounded with power grids and also vulnerable to disasters [15], resulting in simultaneous damage. For instance, the 2008 ice and snow disaster in southern China damaged both cyber and physical infrastructures of DSs, causing widespread outages [16]. DS automation-based restoration relies heavily on the supporting communication network. Disruptions in communication between operation centers and terminal devices impair DS observability and controllability, affecting swift restoration. Typically, cyber failures add complexity to decision-making and necessitate the co-optimization of physical and cyber recovery efforts [17].

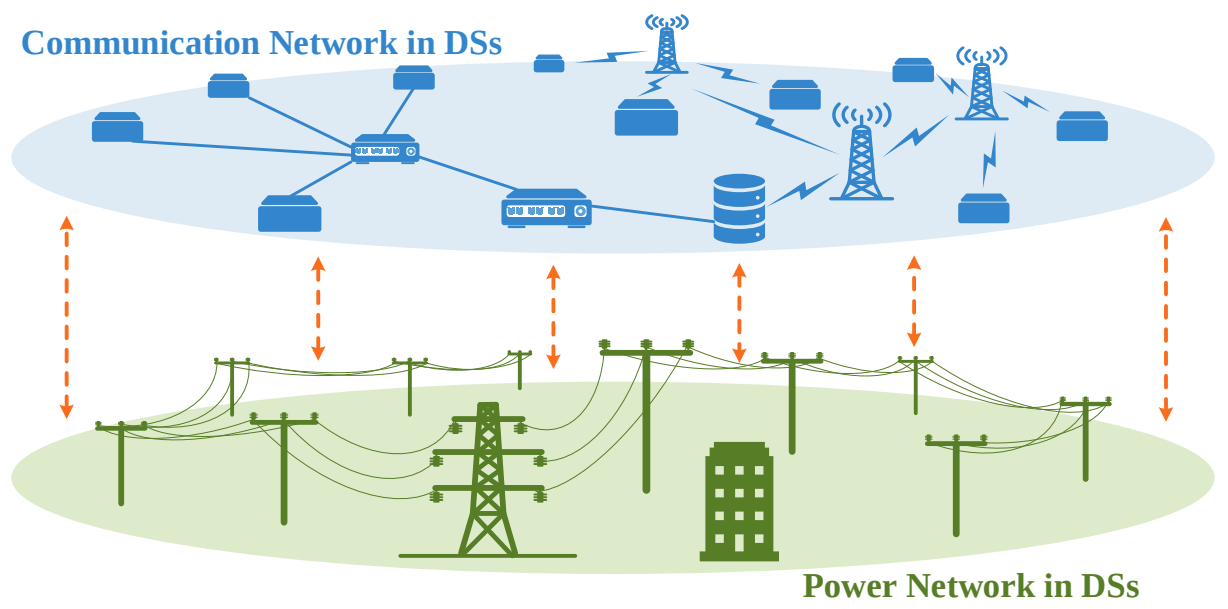

**Fig. 1.** The interdependency of the communication and power networks in DS.

Efforts to address communication failures in DSs, including mobile ad hoc network restoration schemes [17], 5G technology [19], self-healing technologies [20], and cooperative device-to-device communication [21], have been significant. However, these methods primarily focus on restoring communication networks rather than loads. Communication network failures often present challenges such as limited resources (e.g., bandwidth and routing paths), hindering full network restoration. When only a portion of the communication network is recoverable, decision-making in communication recovery significantly affects load recovery effectiveness. Consequently, applying existing communication recovery research directly is not entirely effective in DS restoration.

Addressing this issue involves considering the interdependence of communication and load recovery in DS restoration. The primary aim should be maximizing the load pickup, rather than merely enhancing direct communication metrics (e.g., connectivity or node recovery as noted in [20] and [21]). This necessitates a flexible and reliable communication network with superior control capabilities to sync with DS automation. In this context, software-defined networking (SDN) technology offers significant benefits. Unlike traditional networks where data and control planes are bundled within switches, SDN separates these components [22]. This separation turns network switches into mere forwarding devices, with data paths easily alterable via flow tables from controllers, enhancing link flexibility and control. The SDN enables the integration of various communication technologies via dynamic software configurations [23]. Research indicates that SDN can swiftly identify and isolate faulty links, and restore data interactions [22], providing scalable, robust communication for DS [24]. While SDN's potential to boost power system communication resilience is evident, its application in cyber-physical integrated DS restoration has been largely overlooked.

To address the aforementioned challenges, we introduce an SDN-based DS restoration approach. Initially, we modify traditional DS communication networks by adding alternative backup communication links. Subsequently, SDN technology is implemented in these networks to facilitate centrally managed data routing. Communication resilience is enhanced by SDN controllers issuing routing tables that control data routing paths. Furthermore, we propose incorporating communication demand constraints for remote line control, integrating communication facility statuses into the power load recovery process. We then develop a cyber-physical integrated DS restoration model, designed to optimize load recovery. Moreover, a cyclic algorithm is proposed to effectively utilize scarce communication resources, enabling controlled, multi-step maximum load recovery in DSs. The paper's contributions are outlined as follows.

- We introduce an SDN-based method for DS communication recovery, which is designed to centrally recover failed links under limited resource conditions.
- We model the dependency of power line control on communication as linear constraints in cyber and physical failures and address communication and load recovery problems via the MILP method.
- We develop a multi-step, cyber-physical integrated algorithm aimed at maximizing the pickup load in DSs with constrained communication resources.
- We conduct evaluation experiments to demonstrate the validity of the proposed model for improving the DS restoration effect.

The remainder of this paper is organized as follows: Section II introduces an SDN-based communication network recovery scheme and outlines the cyber-physical integrated DS restoration framework. Section III details the formulation of the DS restoration model, incorporating cyber-physical interdependencies. Section IV presents numerical results from IEEE 33-node and IEEE 123-node test feeders, demonstrating the efficacy of the proposed model. Section V draws a conclusion.

## II. Cyber-Physical Integrated Restoration Framework

We categorize post-disaster cyber-physical DS restoration into two main parts: communication recovery and load recovery. Load recovery in DS involves operating power lines — opening and closing them to supply power from DGs — reliant on automated switch controls. However, such remote controls hinge on a functional communication network. Essentially, communication restoration between operation centers and terminal devices (such as feeder terminal units and distribution terminal units) underpins DS automation control. Leveraging the SDN-based network's data routing capabilities allows for the re-establishment of this crucial communication, thus aiding load recovery. Building on this, we propose a unified cooperative recovery model that simultaneously addresses communication and power recovery needs with the ultimate goal of maximum pickup load.

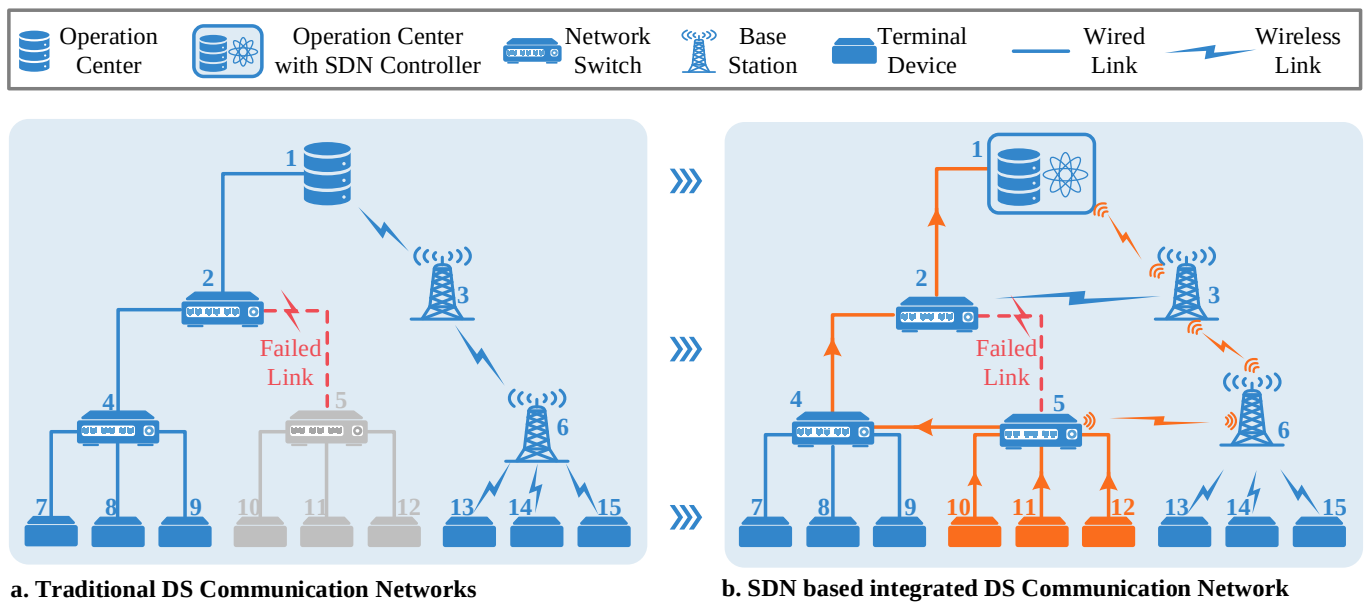

**Fig. 2.** Example of communication network modification and recovery.

### *A. Communication Network Modification*

In DSs, communication is usually facilitated by separate networks utilizing dedicated cables, fibers, and devices [25], as depicted in Fig. 2(a). However, these networks often have rigid structures and lack flexibility in resource allocation and failure recovery. To enhance synergistic power and communication restoration post-disaster (Fig. 2(b)), we modify DS communication networks in two key ways: adding looped links and integrating SDN technology.

1) *Alternate Looped Links:* We loop the forwarding facilities (e.g., network switches and base stations) according to their proximity in terms of spatial location, offering alternate data transmission routes.

2) *SDN Technology Integration:* We equip the operation center with SDN controllers and upgrade network communication facilities to support SDN. This shift to SDN allows data transmission paths to be dynamically controlled by the SDN controller's flow tables.

Subsequent to these modifications, in the event of communication failures, looped links serve as alternate channels for restoring failed connections. With SDN integration, the operation center can use routing tables to adaptively alter data paths for communication restoration, aligning with the specific communication needs of DSs.

Fig. 2 illustrates the described communication recovery function. If the connection between network switches 2 and 5 in Fig. 2(a) fails, all connected terminal devices of these switches lose communication with the operation center, disabling its monitoring and control of the relevant electric utility. Such failures diminish the load recovery effectiveness of DSs. In contrast, as shown in Fig. 2(b), the addition of looped links provides alternative paths, allowing the controller to restore communication in the same failure scenario. Connections can be re-established by rerouting data through adjacent forwarding devices (network switch 4 and base station 6) to the operation center. This approach enables the centralized control recovery capability for DS communication networks.

### *B. Load and Communication Integrated Recovery*

DS automation provides DSs with the capability for rapid microgrid formation using DGs. However, this control is contingent upon communication between terminal devices and the operation center. Based on the above centralized communication recovery capability enabled by communication network modifications, we integrate the DS load recovery model with the communication network recovery model through line controllable constraints. Solving this integrated model computationally derives optimal data routing and line control schemes for load recovery. Following these schemes, SDN controllers issue routing tables to forwarding devices to restore communication connection between terminal devices and the operation center. Subsequently, microgrids are formed by controlling DS line switches. This communication restoration augments DS automation's load recovery effectiveness.

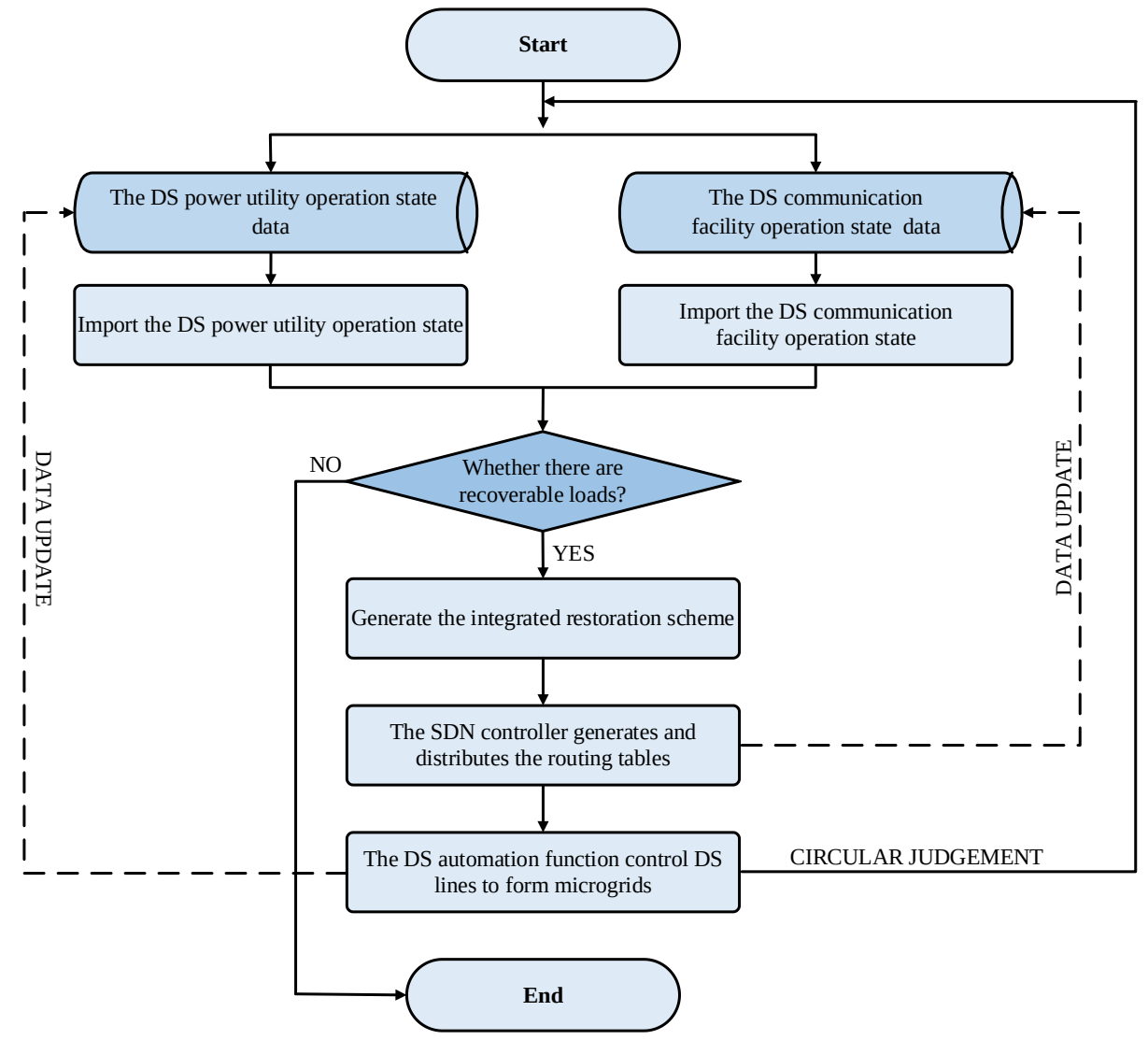

**Fig. 3.** Flowchart of the integrated restoration method for communication and power networks in DSs.

Additionally, in widespread DS failures with constrained communication resources, single-stage restoration may not achieve optimal restoration. Thus, we introduce a cyclic algorithm to assess recoverability after each restorative step, enhancing the restoration effect. This method is detailed in the flowchart shown in Fig. 3.

This methodology enables swift centralized control recovery in cyber-physical faults. Consequently, this approach enhances the DS resilience and restoration speed, while mitigating the need for on-site maintenance—a common requirement in the face of communication failures in traditional restoration methods that necessitate manual site visits for repairs and control during disasters.

## III. MATHEMATICAL FORMULATION

This section describes the cyber-physical integrated restoration model and the corresponding multi-step restoration algorithm. For clarity, 'nodes' and 'links' refer to communication nodes and links, respectively, while 'buses' and 'lines' denote DS buses and lines, respectively.

### *A. Power Load Recovery Model Formulation*

Physically, a DS can be modeled as an undirected graph $\mathcal{G}^p = [\mathcal{N}^p, \mathcal{L}^p]$. $\mathcal{N}^p$ is the set of $\Phi$ buses, indexed by $i$ and $j$.

$\mathcal{L}^p$ is the set of lines, indexed by $k = (i, j)$. Additionally, we denote all buses connected to power sources in a set $\mathcal{S}^p \subset \mathcal{N}^p$.

*1) DS Operational Constraints (DOC):* Since buses and lines may be damaged randomly in extreme events, binary parameters $\chi_i^{PN}$ and $\chi_k^{PL}$ are used to indicate the equipment states of bus $i$ and line $k$, respectively, where "1" indicates normal operation and "0" indicates malfunctions. When a bus or a line fails due to permanent fault, it must be isolated from DSs before manual maintenance is performed. Therefore, line $k = (i,\ j) \in \mathcal{L}$ cannot be closed unless the line and the both end buses function normally, which can be written as

$$b_k \le (\chi_k^{PL} + \chi_i^{PN} + \chi_j^{PN})/3, \forall k = (i,\ j) \in \mathcal{L}, \quad (1)$$

where variable $b_k$ denotes the opening and closing states of line $k$, with "1" indicating closing and "0" indicating opening.

Considering the utility characteristics of DGs and lines, the active and reactive power of DGs and lines should not exceed their upper power limits, which can be described by the following four inequality constraints:

$$0 \le p_i^G \le P_i^{G_max}, \forall i \in \mathcal{S}^p, \quad (2)$$

$$0 \le q_i^G \le Q_i^{G_max}, \forall i \in \mathcal{S}^p, \quad (3)$$

$$-P_k^{L_max} \le p_k^L \le P_k^{L_max}, \forall k \in \mathcal{L}^p, \quad (4)$$

$$-Q_k^{L_max} \le q_k^L \le Q_k^{L_max}, \forall k \in \mathcal{L}^p, \quad (5)$$

where $p_i^G$ and $q_i^G$ denote the active and reactive power, respectively, injected into bus $i$ by the generator connected to it; $P_i^{G_max}$ and $Q_i^{G_max}$ are the maximum active and reactive power limits, respectively, of the generator at bus $i$; $p_k^L$ and $q_k^L$ are the active and reactive power, respectively, flowing from bus $i$ to bus $j$ along line $k = (i,\ j)$; and $P_k^{L_max}$ and $Q_k^{L_max}$ are the maximum active and reactive power limits, respectively, of line $k$.

It is common to have an automated switch between a bus and its load for controlling the load recovery. The binary variable $b_i^{Load}$ denotes the opening and closing states of the automated switch between bus $i$ and its load, with "1" specifies closing and "0" specifies opening. In cases where buses lack load switches, the default state is closed. Both active and reactive power flow only through closed lines, and the sum of respective flows from a bus must be equal to zero. This is represented by the following constraints:

$$-M \cdot b_k \le p_k^L \le M \cdot b_k, \forall k \in \mathcal{L}^p, \quad (6)$$

$$-M \cdot b_k \le q_k^L \le M \cdot b_k, \forall k \in \mathcal{L}^p, \quad (7)$$

$$\textstyle\sum_{k \in \mathcal{F}_i} \mu_i^k \cdot p_k^L + b_i^{Load} \cdot P_i^{Load} = 0, \forall i \in \mathcal{N}^p \backslash \mathcal{S}^p \quad (8)$$

$$\textstyle\sum_{k \in \mathcal{F}_i} \mu_i^k \cdot q_k^L + b_i^{Load} \cdot Q_i^{Load} = 0, \forall i \in \mathcal{N}^p \backslash \mathcal{S}^p \quad (9)$$

$$\textstyle\sum_{k \in \mathcal{F}_i} \mu_i^k \cdot p_k^L + b_i^{Load} \cdot Q_i^{Load} - p_i^G = 0, \forall i \in \mathcal{S}^p, \quad (10)$$

$$\textstyle\sum_{k \in \mathcal{F}_i} \mu_i^k \cdot q_k^L + b_i^{Load} \cdot Q_i^{Load} - q_i^G = 0, \forall i \in \mathcal{S}^p. \quad (11)$$

where a very large number $M$ is used to express the conditional statement and $\mathcal{F}_i$ is the set of lines connected to bus $i$. Specifically, the direction of power flow on a line $k = (i,\ j)$ is defined as from bus $i$ along line $k$ to bus $j$ which can be expressed by parameters $\mu_i^k = 1$ and $\mu_j^k = -1$.

Regarding to the DS operation, the DG bus voltage is set as the reference voltage $V_a$. Additionally, voltages at all buses must adhere to voltage-drop constraints within the DS voltage limits and tolerance $\delta$. These constraints are delineated as follows:

$$v_i = V_a, \forall i \in \mathcal{S}^p, \quad (12)$$

$$(1-\delta) \cdot V_a \le v_i \le (1+\delta) \cdot V_a,\ \forall i \in \mathcal{N}^p, \quad (13)$$

$$v_j + (1-b_k) \cdot (-M) \le v_i - D_{i,k} \cdot (P_k^L \cdot r_k + Q_k^L \cdot x_k)/V_a \le v_j + (1-b_k) \cdot M, \forall k = (i,j) \in \mathcal{L}^p, \quad (14)$$

where variable $v_i$ denotes the voltage of bus $i$; parameters $r_k$ and $x_k$ denote the equivalent lumped resistance and reactance of line $k$, respectively.

*2) DS Connectivity Constraints (DCC):* Demonstrating the efficiency of the connectivity constraints of subgraphs in solving DS restoration [13], a single commodity flow method is applied to form multiple microgrids as follows: All commodities circulate within the same DS topology; a microgrid must have only one DG in accordance with the DS prohibited circulation restriction; a bus must meet its unit-load demands to be connected to DGs; and a unit-load satisfied bus means that there exists one path from that bus to the DGs.

Note that power can only flow through closed lines. If a line is closed, then the electrified states of the buses at both ends should be the same. Moreover, the load switch only can be closed at an energized bus to provide power to its load. Then, the constraints can be expressed as

$$-M \cdot b_k \le f_k^L \le M \cdot b_k, \forall k \in \mathcal{L}^p, \quad (15)$$

$$(1-b_k) \cdot (-M) + f_j^N \le f_i^N \le (1-b_k) \cdot M + f_j^N, \forall k = (i,j) \in \mathcal{L}^p, \quad (16)$$

$$b_i^{Load} \le f_i^N, \forall i \in \mathcal{N}^p, \quad (17)$$

where the integer variable $f_k^L$ denotes the amount of commodity flow from bus $i$ to bus $j$ on line $k = (i,j)$ and the binary variable $f_i^N$ represents the unit demand of bus $i$, with "1" indicating that the bus is connected to a DG, and "0" indicating otherwise.

Since the sum of power flowing out of a bus must be zero, they are described by the following equality constraints:

$$\textstyle\sum_{k \in \mathcal{F}_i} \mu_i^k \cdot f_k^L + f_i^N = 0, \forall i \in \mathcal{N}^p \backslash \mathcal{S}^p, \quad (18)$$

$$\textstyle\sum_{k \in \mathcal{F}_i} \mu_i^k \cdot f_k^L + f_i^N = f_i^S, \forall i \in \mathcal{S}^p, \quad (19)$$

where $f_i^S$ denotes the amount of commodity injected into bus $i$ by the generator connected to it.

Additionally, the radial topology constraint is introduced to prevent circulating flows and ensure that a microgrid has only one DG, which can be written as

$$\textstyle\sum_{i \in \mathcal{N}^p} f_i^N - \sum_{i \in \mathcal{S}^p} f_i^N = \sum_{k \in \mathcal{L}^p} b_k. \quad (20)$$

### *B. Communication Recovery Model Formulation*

DS communication networks are typically modeled as undirected graphs $\mathcal{G}^c = [\mathcal{N}^c, \mathcal{L}^c]$. $\mathcal{N}^c$ is the set of communication nodes representing network switches, including network switches, base stations, terminal devices, and the operation center, identified by indices $m$ and $n$. $\mathcal{L}^c$ is the set of wired and wireless communication links, indexed by $l = (m, n)$. Communication nodes and links may fail due to extreme events. To represent the operational status of nodes and links, binary parameters $\chi_m^{CN}$ and $\chi_l^{CL}$ are used, with "1" indicating normal operation and "0" indicating a malfunction. Terminal devices are presumed to be capable of self-testing and reporting their status, while network switches determine link states using the link layer discovery protocol (LLDP) [26]. This enables the operation center to ascertain the working states of nodes and links.

*1) Data Flow Constraints (DFC):* To allocate bandwidth resources in a manner, $c_m^N = (h_1^m, \dots\dots, h_\Phi^m)$ and $c_l^L = (h_1^l \dots\dots, h_\Phi^l)$ are two vectors consisting of binary variables, with $h_i^m$ and $h_i^l$ being binary variables indicating whether the data of terminal device $i$ flow through link $k$ and node $i$, respectively. The data can only pass through normally functioning links and nodes, that is,

$$-\chi_l^{CL} \le h_i^l \le \chi_l^{CL}, \forall l \in \mathcal{L}^C, \forall i \in \mathcal{N}^p, \quad (21)$$

$$-\chi_m^{CN} \le h_i^m \le \chi_m^{CN}, \forall m \in \mathcal{N}^C, \forall i \in \mathcal{N}^p. \quad (22)$$

For the operation center node, the data to be received should consist of all the data on the connected communication links. For forwarding nodes, there should be two procedures for sending and receiving identical data. For terminal nodes, there should be only one data transmission process. The above characteristics can be translated into the following three linear equality constraints:

$$\textstyle\sum_{l \in \mathcal{H}_o} c_l^L = c_o^N, \quad (23)$$

$$\textstyle\sum_{l \in \mathcal{H}_m} c_l^L = 2c_m^N, \forall m \in \mathcal{N}^c \backslash (\mathcal{N}^p \cup o), \quad (24)$$

$$\textstyle\sum_{l \in \mathcal{H}_m} c_l^L = c_m^N = E_\Phi(m,:) \cdot s_m, \forall m \in \mathcal{N}^p, \quad (25)$$

where node $o$ represents the operation center; $\mathcal{H}_i$ is the set of links connected to node $i$; and $E_\Phi(m,:)$ denotes row $m$ of the $\Phi$-dimensional identity matrix.

Furthermore, the communication state of a terminal device node is determined by whether the operation center can receive the data from this terminal device. Only normally functioning terminal devices can communicate with the operation center, which is described as

$$s_m = h_m^o, \forall m \in \mathcal{N}^p, \quad (26)$$

$$s_m \le \chi_m^{CN}, \forall m \in \mathcal{N}^p, \quad (27)$$

where binary $s_m$ is the communication state of terminal device node $m$ to operation center node $o$, with "1" indicating connected and "0" indicating disconnected.

*2) Bandwidth and Delay Constraints (BDC):* The bandwidth consumption of a node or link is equal to the sum of the bandwidth consumption of the data flowing through it and cannot exceed its bandwidth capacity. These constraints can be expressed as

$$d_m^N = \textstyle\sum_{i \in \mathcal{N}^p} w_i \cdot h_i^m, \forall m \in \mathcal{N}^C, \quad (28)$$

$$d_m^N \le \lambda_m^{max}, \forall m \in \mathcal{N}^C, \quad (29)$$

$$d_l^L = \textstyle\sum_{i \in \mathcal{N}^p} w_i \cdot h_i^l, \forall l \in \mathcal{L}^c, \quad (30)$$

$$d_l^L \le \lambda_l^{max}, \forall l \in \mathcal{L}^c, \quad (31)$$

where $d_m^N$ and $d_l^L$ denote the bandwidths consumed by the data through node $m$ and link $l$, respectively; $w_i$ is the required communication bandwidth for terminal device $i$; and $\lambda_m^{max}$ and $\lambda_l^{max}$ denote the bandwidth upper limits of node $m$ and link $l$, respectively.

The end-to-end communication delay between a node and the operation center is the sum of the forwarding delays of the forwarding nodes and the propagation delays of the links through which the data are passed. This end-to-end delay cannot exceed the specified delay limitation, that is,

$$e_i = \textstyle\sum_{l \in \mathcal{L}^c} h_i^l \cdot \xi_l^L + \sum_{m \in \mathcal{N}^c \backslash \mathcal{N}^p} h_i^m \cdot \xi_m^N, \forall i \in \mathcal{N}^p, \quad (32)$$

$$e_i \le \tau_i^{max}, \forall i \in \mathcal{N}^p, \quad (33)$$

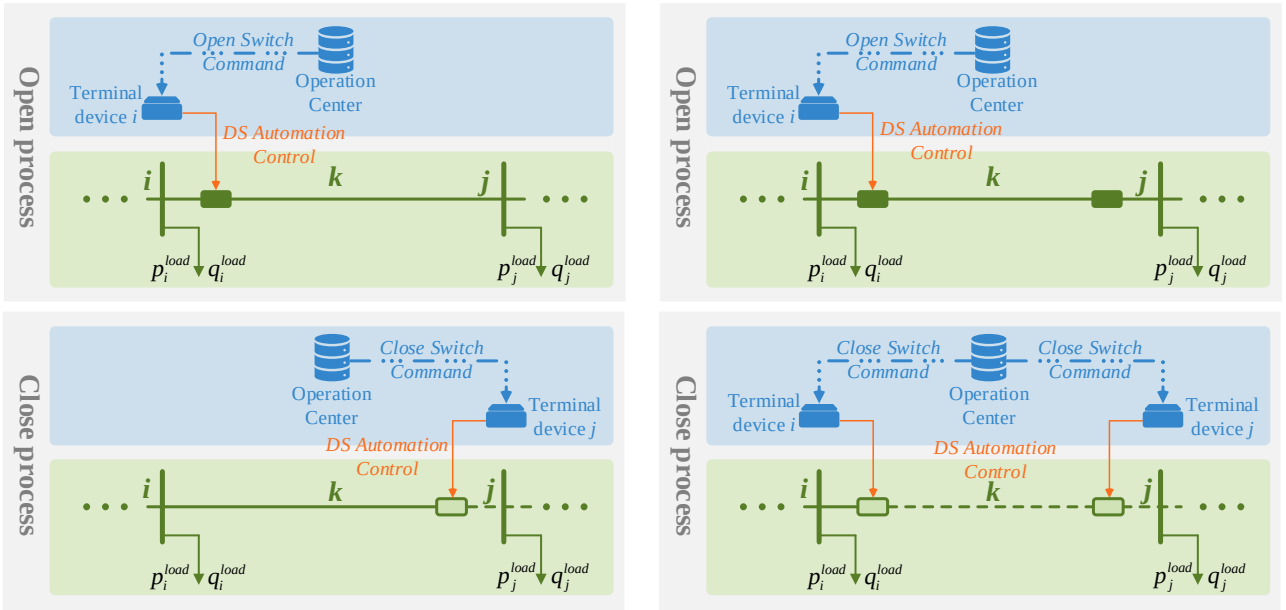

a. Control process for lines equipped with automated switches on one side

b. Control process for lines equipped with automated switches on both sides

**Fig. 4.** Schematic diagram of the control of different categories of DS lines.

where $e_i$ is the end-to-end data delay from terminal device node $i$ to the operation center; $\tau_i^{max}$ denotes the upper limit on the end-to-end delay for its DS service; and $\xi_l^L$ and $\xi_m^N$ denote the propagation delay of link $l$ and the forwarding delay of node $m$, respectively.

### *C. Cyber-Physical Integrated Constraint and the Objective Function*

*1) Line Control Constraints (LCC):* Most DS lines have automated switches near the buses, monitored and controlled by terminal devices. Line remote control is achieved by transmitting control commands from the operation center to the corresponding terminal device via the communication network. Some DS lines are equipped with automated switches on only one side, whereas crucial lines might have switches on both end sides (e.g., DS contact lines).

As depicted in Fig. 3, lines in various initial operational states have distinct communication needs. Hence, we categorize line control communication requirements into six scenarios in Table I and simplify them into an inequality constraint as:

$$\varpi_k^L - (\varrho_i^k \cdot s_i + \varrho_j^k \cdot s_j) \le b_k \le \varpi_k^L + (\varrho_i^k \cdot s_i + \varrho_j^k \cdot s_j) / (\varrho_i^k + \varrho_j^k), \forall k = (i,j) \in \mathcal{L}^p, \quad (34)$$

where the binary parameter $\varrho_i^k$ denotes whether a switch is installed on the node $i$'s side of line $k$.

TABLE I

INFLUENCE OF LINE OPERATIONAL STATE ON COMMUNICATION REQUIREMENT IN LINE CONTROL

| Line $k$ state | The presence of switches at sides of buses and their states | | | | Control state | Communications requirements | |
|---|---|---|---|---|---|---|---|
| | Bus $i$ | | Bus $j$ | | | Bus $i$ | Bus $j$ |
| Closed | Has | Closed | No | / | Open | $s_i = 1$ | / |
| | No | / | Has | Closed | | / | $s_j = 1$ |
| | Has | Closed | Has | Closed | | $s_i + s_j \ge 1$ | |
| Open | Has | Open | No | / | Closed | $s_i = 1$ | / |
| | No | / | Has | Open | | / | $s_j = 1$ |
| | Has | Open | Has | Open | | $s_i = 1$ | $s_j = 1$ |

For the safe operation of the DS, it is necessary for the operation center to observe both buses at the ends of a line before closing it. Thus, constraint (34) can be redefined as

$$\varpi_k^L - (\varrho_i^k \cdot s_i + \varrho_j^k \cdot s_j) \le b_k \le \varpi_k^L + 0.5 \cdot (s_i + s_j), \forall k = (i,j) \in \mathcal{L}^p. \quad (35)$$

Notably, lines lacking automated switches are non-controllable and thus exempt from the aforementioned constraints.

With (35), the communication state of terminal devices in communication networks and line control for recovery in power networks are integrated. This integration allows for the simultaneous determination of data routing paths and line controls for load recovery in a cyber-physical integrated restoration model, using a single objective function. The outcomes, including device and link routing variables ($c_m^N$ and $c_l^L$), along with line control variables ($b_k$) from this function, define the data routing for communication recovery and the approach for microgrid formation.

*2) Objective Function for Cyber-Physical Integrated Restoration (OFCPIR):* The value of loads may differ by their category [27]. The objective function for this cyber-physical integrated restoration is the maximum value of the restored load with the minimum end-to-end communication delay under communication resource constraints. The objective function is formulated as follows:

$$max \sum_{i\in\mathcal{N}} p_i^{Load} \cdot b_i^{Load} \cdot \phi_i + min \sum_{i\in\mathcal{N}} \varepsilon \cdot e_i, \qquad (36)$$

$$s.t.\ \text{DOC}: (1)-(14),$$
$$\text{DCC}: (15)-(20),$$
$$\text{DFC}: (21)-(27),$$
$$\text{BDC}: (28)-(33),$$
$$\text{LCC}: (35),$$

where $\phi_i$ is the weight parameter of load in bus $i$ and $\varepsilon$ is a very small parameter used in solving for the smallest end-to-end delay of terminal devices.

Thus, the whole model can be formulated as an MILP problem that can be solved using off-the-shelf solvers.

*3) Cyclic Algorithm:* Given the extensive fault conditions and limited communication resources, which restrict the number of automated switches controllable in a single restoration step, fully restoring all recoverable loads in one process is challenging. Consequently, we introduce a cyclic algorithm to develop a multi-step restoration scheme via iterative assessments. The algorithm is detailed as follows:

**Algorithm 1** Multi-Step Restoration Algorithm

**Input:** The DS topology $\mathcal{G}^p = [\mathcal{N}^p, \mathcal{L}^p]$ along with the equipment states, operation states, and parameters;
DS communication network topology $\mathcal{G}^c = [\mathcal{N}^c, \mathcal{L}^c]$ along with the equipment states, operation states, and parameters;
The cyclic parameter $\xi = 1$;

```
while ξ = 1
    Compute objective function (OFCPIR) of the cyber-physical
    integrated restoration model based on the states and parameters;
    if OFCPIR has a restoration solution
        Records the link routing scheme (c_m^N, c_l^L), load recovery
        control scheme b_k;
        Update the DS and communication network states;
    else
        ξ = 0;
    end if
end while
```

**Output:** the whole cyber-physical integrated restoration scheme

Using the above algorithm, a multi-stage communication and load co-recovery scheme can be developed. This scheme maximizes the load recovery by sequentially controlling the data routing within the communication network and line splitting in the power network.

## IV. Numerical Results

This section presents an evaluation of the proposed MILP model using IEEE 33-node and IEEE 123-node test feeders [28] under dual cyber-physical fault scenarios. The model's effectiveness is validated by comparing its results with those of two benchmark methods:

*a. Only Load Recovery (OLR) Algorithm:* This algorithm focuses solely on load recovery, conforming to BCC constraints, and aims to maximize the recovered load in line with the objective of [13].

*b. Separated Cyber and Load Recovery (SCLR) Algorithm:* This is a two-step algorithm that addresses both communication and load recovery. Initially, it focuses on maximizing the number of recovered communication nodes using network resources, in line with [13]. Subsequently, it undertakes load recovery, considering BCC constraints and aiming to maximize recovered load.

*c. Integrated Cyber and Load Recovery (ICLR) Algorithm:* This is the proposed cyber-physical integrated recovery approach in this paper. It combines communication and load recovery, optimizing the communication and load recovery while adhering to their constraints.

The calculations for the three algorithms are implemented on a PC with an Intel Core i7 CPU @2.90 GHz and 32 GB of memory. The YALMIP toolbox in MATLAB 2021b with Gurobi 9.1.2 is utilized to solve the MILP problems, and the gap for the MILP problems is set to 0.0001. Mininet, a tool for simulating SDN-enabled communication networks [24], was employed. The data routing schemes for SDN communication networks, as computed in our model, were implemented and tested on Mininet 2.3.1 to validate their effectiveness.

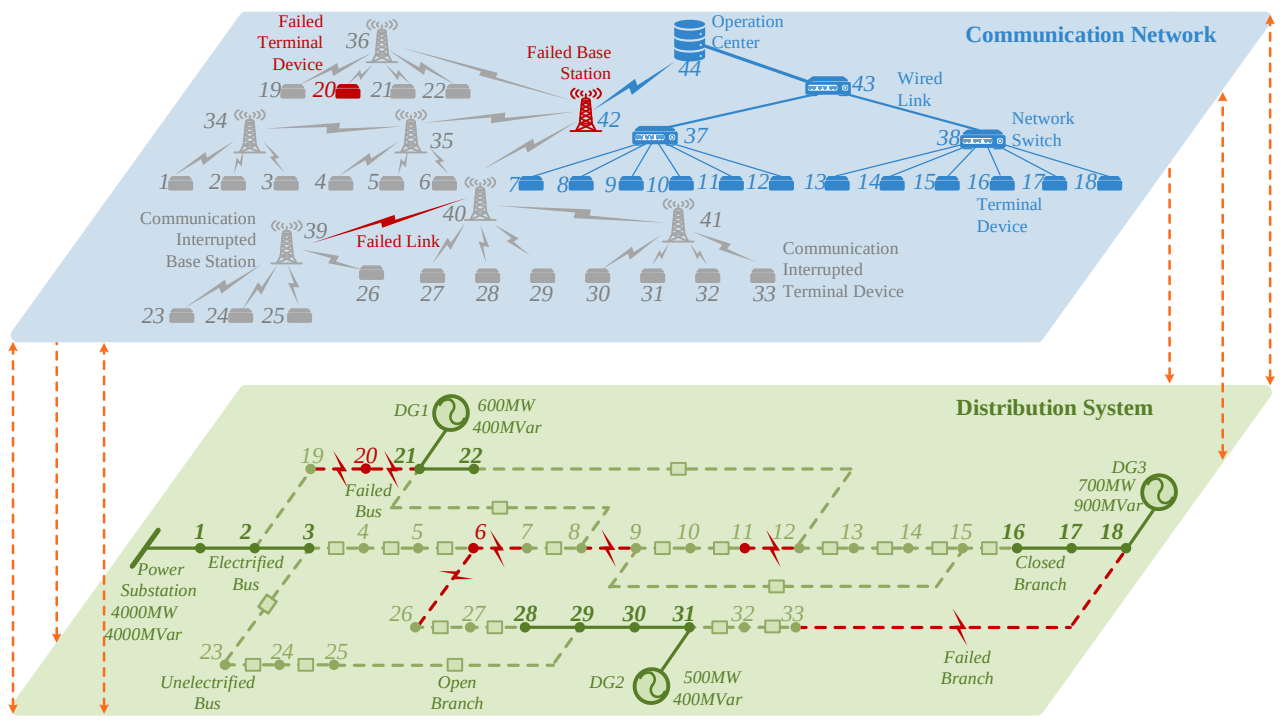

**Fig. 5.** IEEE 33-node test feeder case.

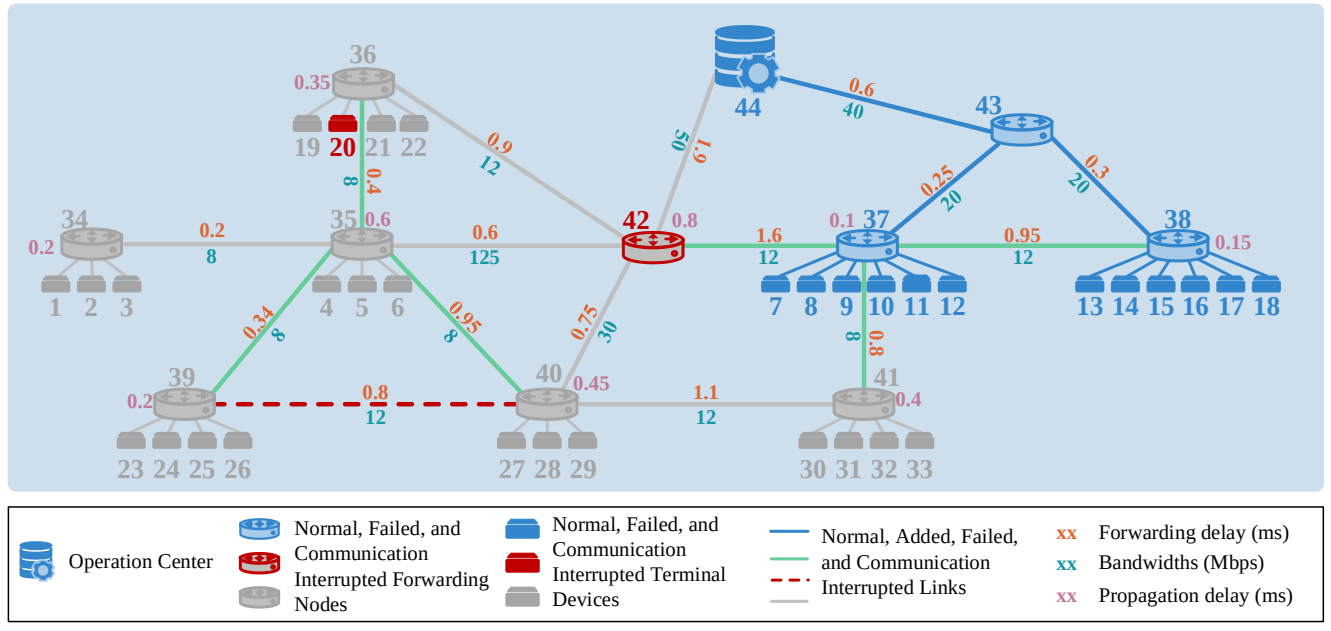

**Fig. 6.** IEEE 33-node test feeder communication network diagram.

### A. IEEE 33-node Test Feeder Case

An IEEE 33-node test feeder at a nominal voltage of 12.66 kV with its 44-node communication network is built as shown in Fig. 5. Three DGs are installed at three nodes 18, 21, and 31.

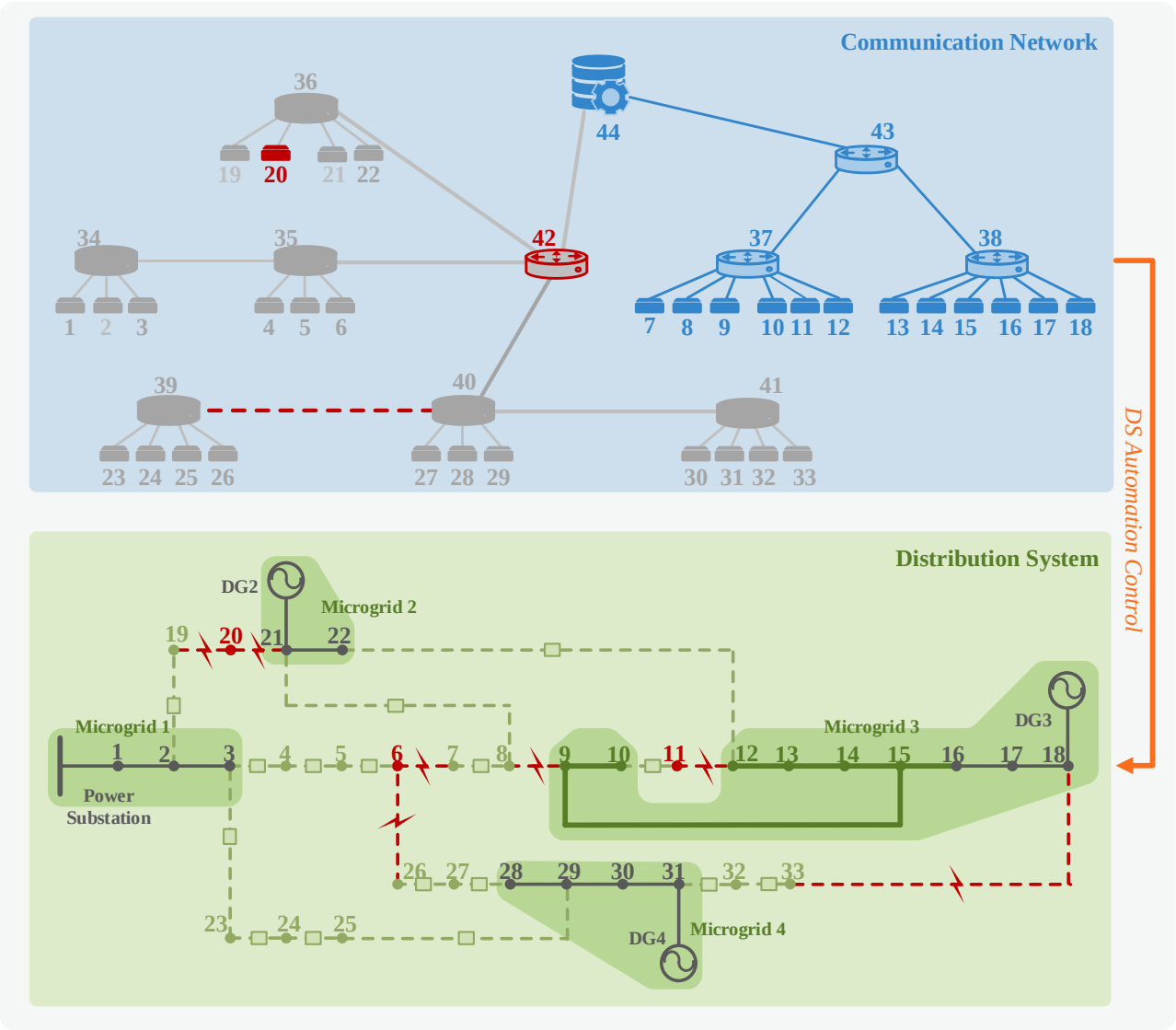

**Fig. 7.** IEEE 33-node test feeder recovery results for the OLR algorithm.

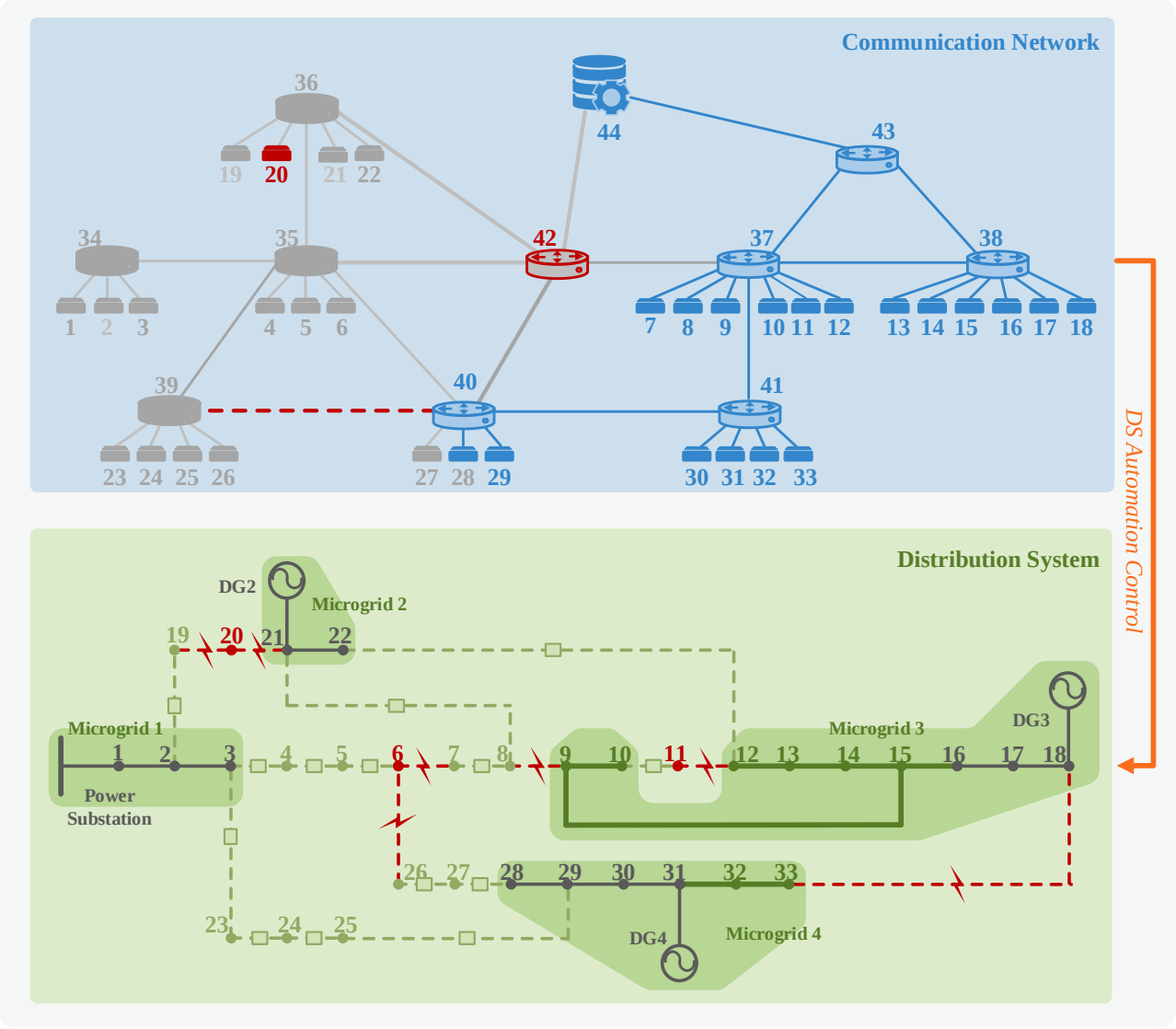

**Fig. 8.** IEEE 33-node test feeder recovery results for the SCLR algorithm.

The communication network employs both wired and wireless methods. The parameters for the real and reactive power capacities of the power sources are also illustrated in Fig. 5. Note that in practice, how the values of the weights are chosen is a complicated problem but is not the focus of this paper. Therefore, the load weight parameters ($\phi_i$) are assumed to be 1. Since the store-and-forward method is the mainstream method for message forwarding in switches, we assume the forwarding delay rule on this basis. The maximum end-to-end delay $e_x^{CN_max}$ is set to 10 ms according to the communication requirements for DS data transmission in IEC 61850-5 [29], and the bandwidth requirement for a terminal device is assumed to be 2 Mbps [30]. Details on the additional links, as well as the bandwidth and delay parameters for nodes and links, are presented in Fig. 6.

In this case, a disaster impacts this DS, damaging both power and communication infrastructures. Several lines are disrupted and communications between certain terminal devices and the operation center are interrupted due to temporary or permanent faults.

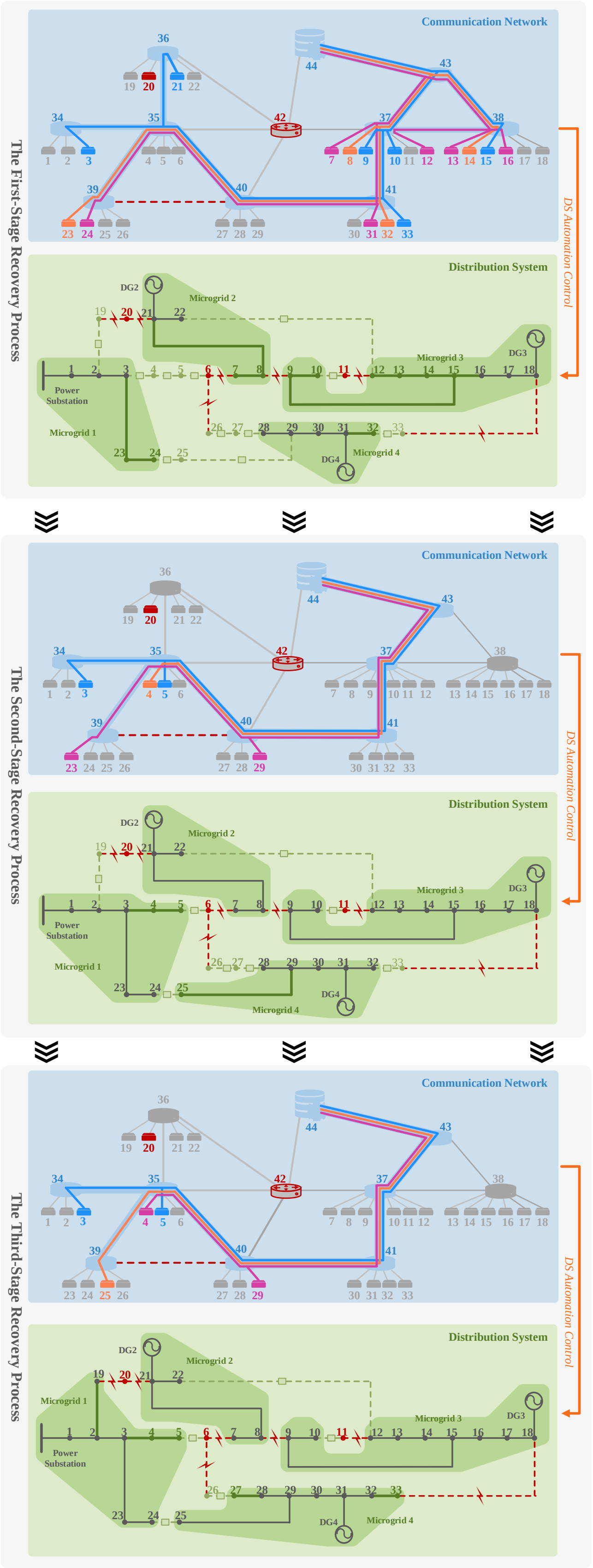

**Fig. 9.** IEEE 33-node test feeder recovery results for the ICLR algorithm.

The three proposed algorithms are applied to compute restoration schemes in this IEEE 33-node test feeder case. The communication and power networks recovered using these algorithms are illustrated in Figs. 7, 8, and 9. For the ICLR algorithm, terminal devices and their data routing paths are categorized into three groups (depicted in three colors) to demonstrate their data routing paths in Fig. 9. Additionally, Fig. 10 displays the computation times, the number of communication nodes, the number of energized buses, and the load recovery for each algorithm. Furthermore, the SDN networks corresponding to the ICLR algorithm were constructed in Mininet. By modifying the routing table with 'dpctl' commands and testing the communication states from terminal devices to the operation center on the Mininet platform, the effectiveness of the recovery scheme is confirmed.

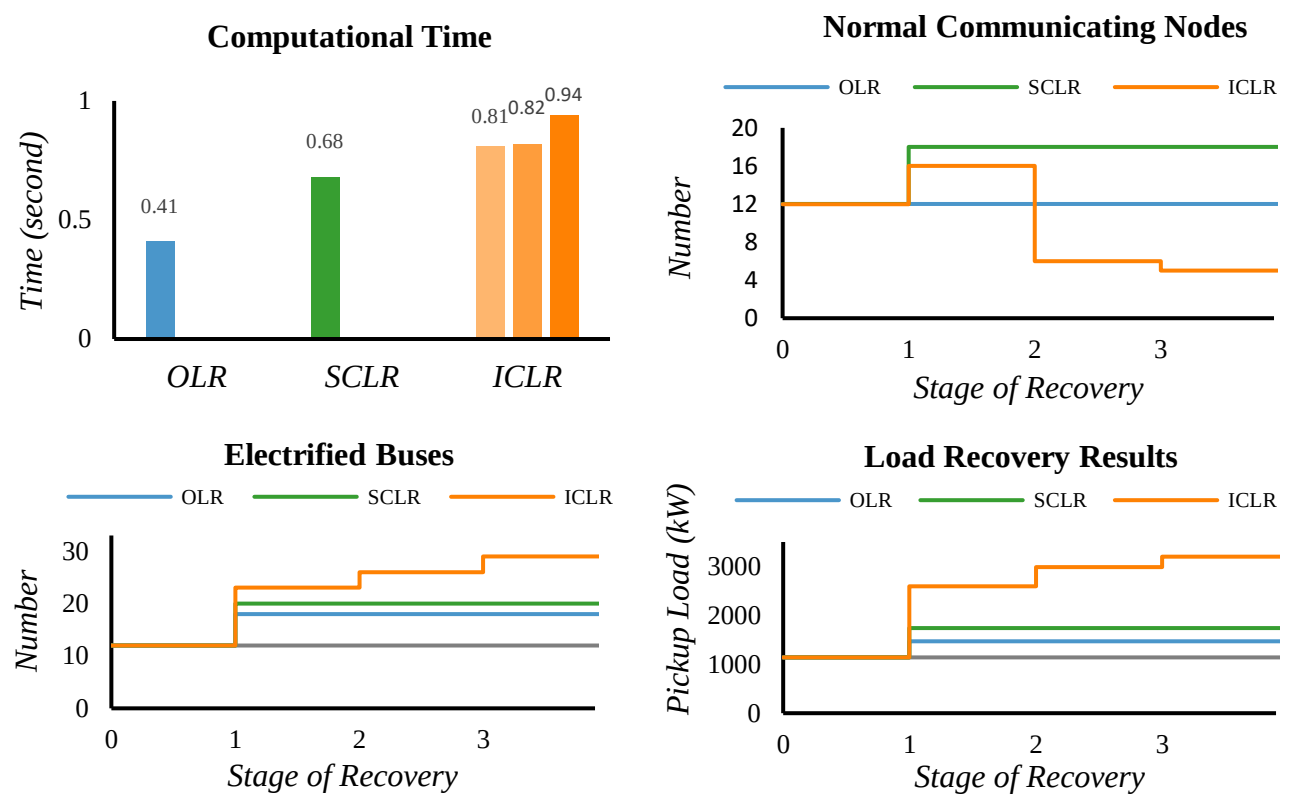

**Fig. 10.** IEEE 33-node test feeder recovery results for the three algorithms.

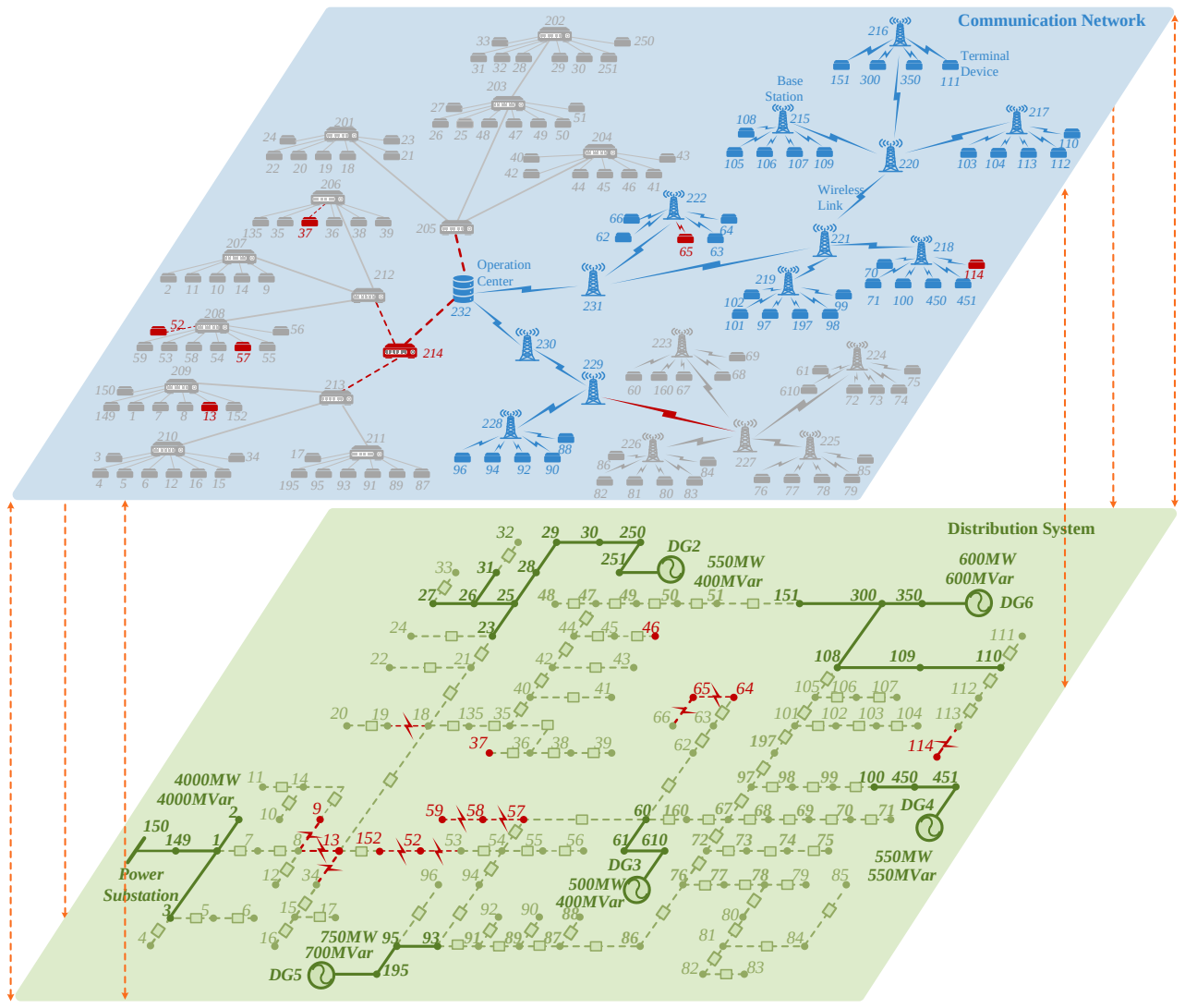

**Fig. 11.** IEEE 123-node test feeder case.

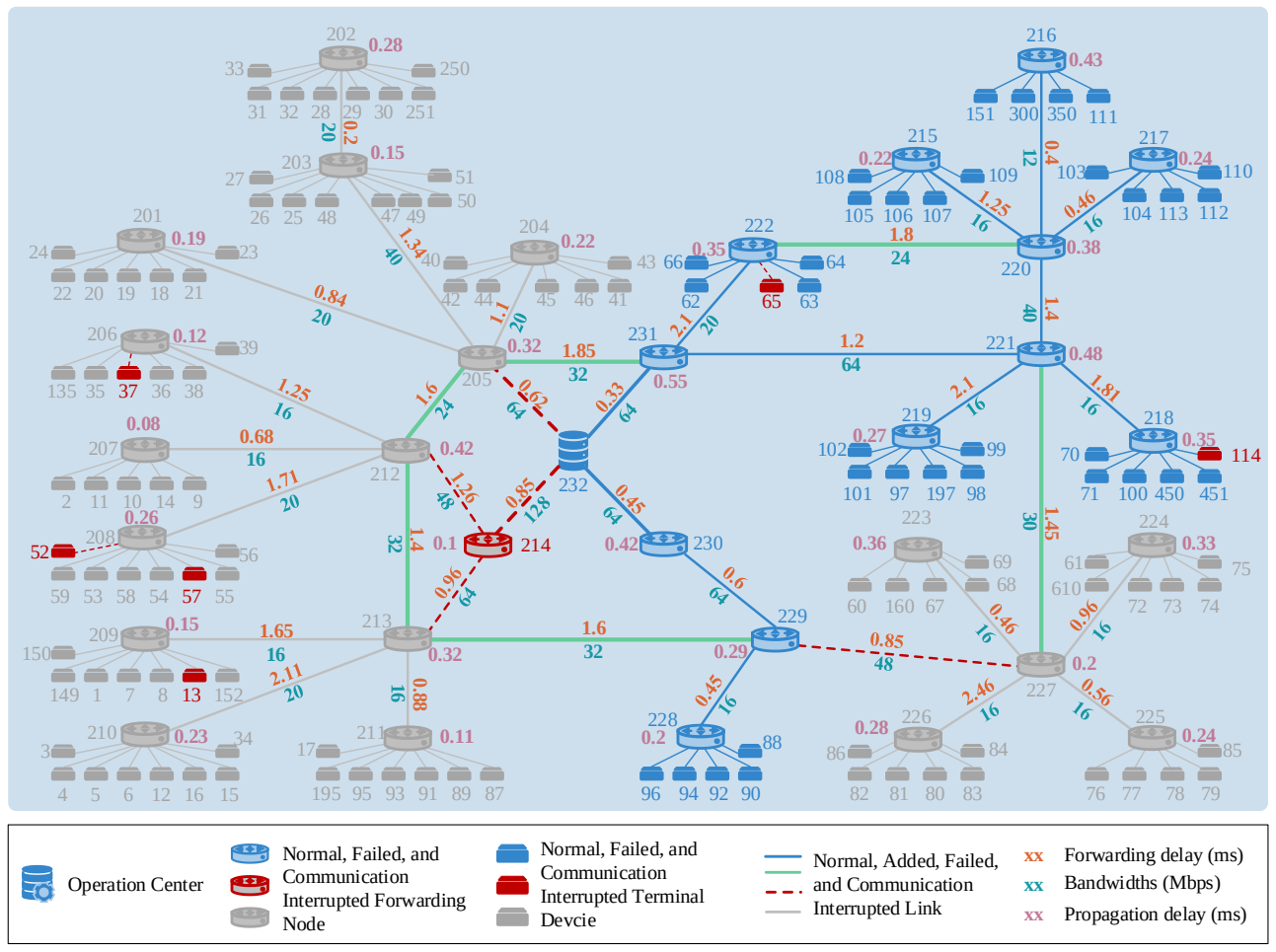

**Fig. 12.** IEEE 123-node test feeder communication network diagram.

### *B. IEEE 123-node Test Feeder Case*

Fig. 11 displays an IEEE 123-node test feeder alongside its 161-node communication network. This DS operates at a nominal voltage of 4.16 kV, with five DGs installed at five buses (195, 251, 350, 451, 610). The power source capacity parameters are detailed in Fig. 11. The added links, along with the node and link parameters for the communication network, are depicted in Fig. 12. Assumptions for load weight, maximum end-to-end delay, and bandwidth requirements for the terminal devices are also set at 1, 10 ms, and 2 Mbps, respectively. The second disaster scenario, illustrated in Fig. 11, resulted in the failure of buses, lines, nodes, and links.

The same three algorithms were employed to calculate the recovery schemes, with the results for the restored communication and power networks displayed in Fig. 13. Additionally, Fig. 14 illustrates the elapsed computation time, number of communication nodes, number of energized buses, and load recovery outcomes for each algorithm. The feasibility of the communication network routing scheme was also confirmed through simulations of the corresponding SDN-based DS communication network on the Mininet platform.

### *C. Discussions of Results*

*1) IEEE 33-node test feeder case:* In the IEEE 33-node test feeder case, the number of electrified buses for the OLR, SCLR, and ICLR algorithms are 18, 20, and 23-26-29 (three-stage restoration) respectively, as shown in Figs. 7, 8, 9, and 10. The load recovery outcomes for these algorithms are 1470kW, 1740kW, and 2590-2990-3200kW (three-stage restoration), respectively, as also indicated in Fig. 10. The results demonstrate that the SCLR and ICLR algorithms, which restore the communication network, recover more load than dose the OLR algorithm. This observation underscores the significant impact of communication network restoration on enhancing the DS load recovery in cyber-physical failures. However, a comparison of load recovery (1740kW and 2590-2990-3200kW) and normally communicating nodal results (18 nodes and 16-6-5 nodes) of SCLR and ICLR algorithms shows that communication restoration schemes that neglect the DS control demands of the DS automation function are less effective. With limited communication resources, the ICLR algorithm restores the network in line with DS control requirements. Utilizing a cyclic algorithm, the ICLR approach effectively leverages limited resources for multi-step recovery to achieve optimal results. Additionally, the computation times for OLR, SCLR, and ICLR algorithms are 0.81s, 0.82s, and 0.91s, respectively, indicating that the ICLR algorithm swiftly provides DSs with a post-disaster restoration solution, ensuring efficient integrated restoration.

*2) IEEE 123-node test feeder case:* Similarly, the SCLR and ICLR algorithms which perform communication recovery, demonstrate superior load recovery performance (45 nodes and

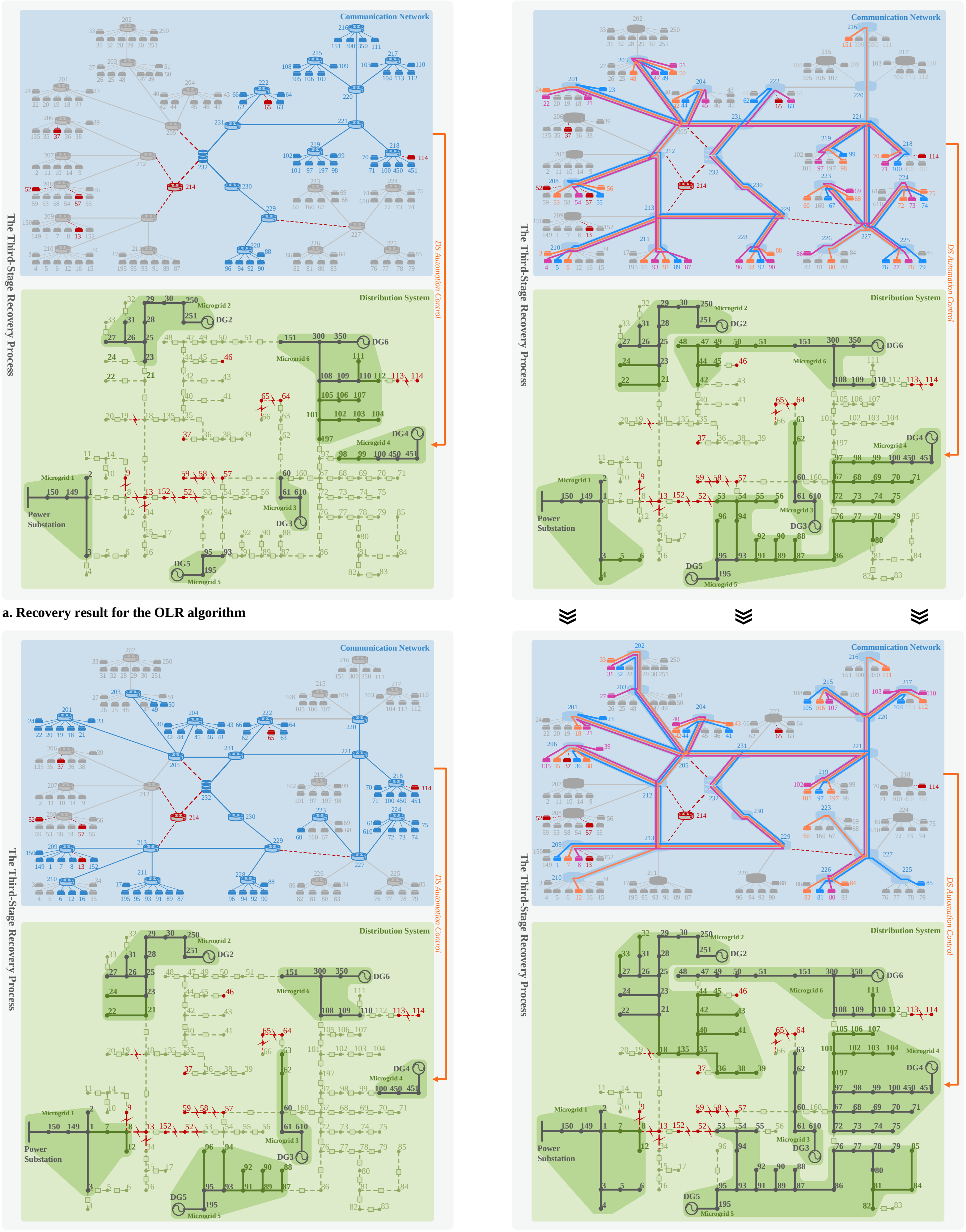

**a. Recovery result for the OLR algorithm**

**b. Recovery result for the SCLR algorithm**

**c. Two-stage recovery results of ICLR algorithm**

**Fig. 13.** IEEE 123-node test feeder recovery results for the three algorithms

740 kW, 76-104 nodes and 2140-2640 kW) compared to the OLR algorithm (42 nodes and 620 kW). Moreover, the ICLR algorithm, which integrates DS automation control requirements, more efficiently utilizes communication resources to attain optimal recovery across multiple stages. The computation times for the OLR, SCLR, and ICLR algorithms are 3.21s, 10.58s, and 15.27-12.09s, respectively, demonstrating the efficiency of the proposed algorithms and affirming the applicability of our model for large-scale DSs. In summary, the results demonstrate that DS restoration is significantly enhanced by implementing cyber-physical integrated restoration schemes.

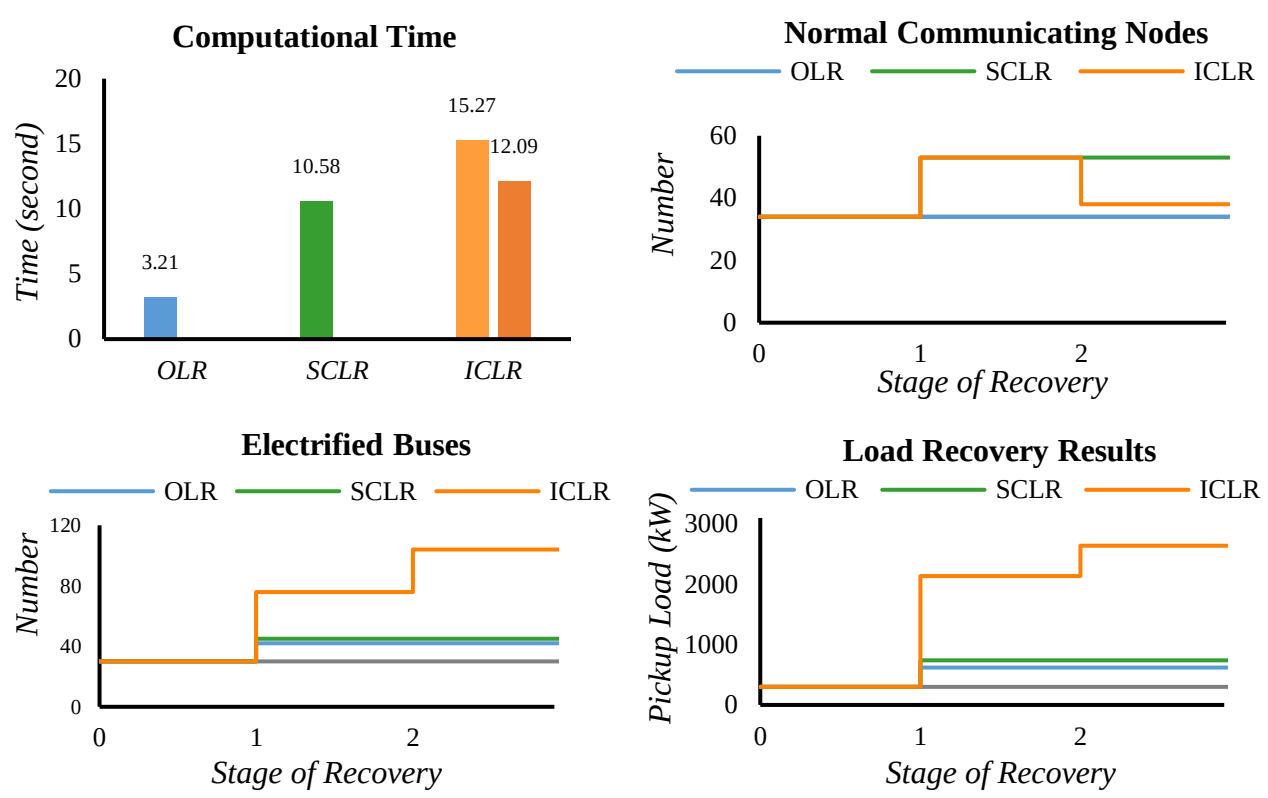

**Fig. 14.** IEEE 123-node test feeder recovery results for the three algorithms.

## V. Conclusion

This paper introduces a cyber-physical integrated restoration model tailored for DS recovery. Leveraging the centralized control routing capabilities of SDN technology, this model effectively restores the communication network, aligning with the needs of DS automation for power restoration and facilitating extensive microgrid formations for load recovery. Additionally, we propose a cyclic algorithm designed to optimize the load recovery via a multi-stage recovery control process. The model's efficacy is confirmed by numerical results from two IEEE test feeders. Future research will focus on a more detailed analysis of the computational complexity and scalability of our method, enhancing the resilience of communication networks, and conducting co-simulations of communication and power network restoration.

## References

[1] L. Xu, Q. Guo, Z. Wang and H. Sun, "Modeling of Time-Delayed Distributed Cyber-Physical Power Systems for Small-Signal Stability Analysis," *IEEE T. Smart Grid*, vol. 12, no. 4, pp. 3425-3437, 2021.

[2] K.S.A. Sedzro, X. Shi, A.J. Lamadrid and L.F. Zuluaga, "A Heuristic Approach to the Post-Disturbance and Stochastic Pre-Disturbance Microgrid Formation Problem," *IEEE T. Smart Grid*, vol. 10, no. 5, pp. 5574-5586, 2018.

[3] S.V. Buldyrev, R. Parshani, G. Paul, H.E. Stanley and S. Havlin, "Catastrophic cascade of failures in interdependent networks," *Nature*, vol. 464, no. 7291, pp. 1025-1028, 2010.

[4] V. Rosato, L. Issacharoff, F. Tiriticco, S. Meloni, S. De Porcellinis and R. Setola, "Modelling interdependent infrastructures using interacting dynamical models," *Int. J. Crit. Infrastruct.*, vol. 4, no. 1-2, pp. 63-79, 2008.

[5] V. Krishnamurthy, A. Kwasinski and L. Duenas-Osorio, "Comparison of power and telecommunications dependencies and interdependencies in the 2011 tohoku and 2010 maule earthquakes," *J. Infrastruct. Syst.*, vol. 22, no. 3, 2016.

[6] S. Mehrdad, S. Mousavian, G. Madraki and Y. Dvorkin, "Cyber-Physical Resilience of Electrical Power Systems Against Malicious Attacks: a Review," *Current Sustainable/Renewable Energy Reports*, vol. 5, no. 1, pp. 14-22, 2018.

[7] C. Chen, J. Wang, F. Qiu and D. Zhao, "Resilient Distribution System by Microgrids Formation after Natural Disasters," *IEEE T. Smart Grid*, vol. 7, no. 2, pp. 958-966, 2016.

[8] D.A. Call, "Changes in Ice Storm Impacts over Time: 1886–2000," *Weather Clim. Soc.*, vol. 2, no. 1, pp. 23-35, 2010.

[9] W. Wang, L. Wang, Y. Miao, C. Cheng and S. Chen, "A survey on the influence of intense rainfall induced by climate warming on operation safety and service life of urban asphalt pavement," *J. Infrastruct. Preserv. and Resil.*, vol. 1, no. 1, 2020.

[10] J. Liu, X. Yang and S. Ren, "Research on the Impact of Heavy Rainfall Flooding on Urban Traffic Network Based on Road Topology: A Case Study of Xi'an City, China," *Land*, vol. 12, no. 7, pp. 1355, 2023.

[11] D. Jin, et al., "Toward a Cyber Resilient and Secure Microgrid Using Software-Defined Networking," *IEEE T. Smart Grid*, vol. 8, no. 5, pp. 2494-2504, 2017.

[12] J. Zhu, Y. Yuan and W. Wang, "An exact microgrid formation model for load restoration in resilient distribution system," *Int. J. Electr. Power Energy Syst.*, vol. 116, 2020.

[13] T. Ding, Y. Lin, G. Li and Z. Bie, "A New Model for Resilient Distribution Systems by Microgrids Formation," *IEEE T. Power Syst.*, vol. 32, no. 5, pp. 4145-4147, 2017.

[14] I.H. Lim, et al., "Security protocols against cyber attacks in the distribution automation system," *IEEE T. Power Deliver.*, vol. 25, no. 1, pp. 448-455, 2010.

[15] Y. Guo, C. Guo and J. Yang, "A tri-level optimization model for power systems defense considering cyber-physical interdependence," *IET Gener. Transm. Distrib.*, vol. 17, no. 7, pp. 1477-1490, 2023.

[16] K. Yaji, H. Homma, T. Aso and M. Watanabe, "Evaluation on flashover voltage property of snow accreted insulators for overhead transmission lines, part II - Flashover characteristics under salt contaminated snowstorm," *IEEE T. Dielect. El. In.*, vol. 21, no. 6, pp. 2559-2567, 2014.

[17] A. Clark and S. Zonouz, "Cyber-physical resilience: Definition and assessment metric," *IEEE T. Smart Grid*, vol. 10, no. 2, pp. 1671-1684, 2019.

[18] B. Chen, M. Ni, C. Yu, and H. Wu, "Research on Optimizing Recovery Strategy of Distribution Network Communications System in Disaster Based on Mobile Ad-Hoc Networks," *The Fourth International Conference on Cyber-Technologies and Cyber-Systems*, Porto, Portugal, 2019, pp. 1293-1298.

[19] A.A.R. Alsaeedy and E.K.P. Chong, "5G and UAVs for Mission-Critical Communications: Swift Network Recovery for Search-and-Rescue Operations," *Mobile Netw. Appl.*, 25, 2063–2081, 2020.

[20] W. Li, Y. Deng, M. Zhang, J. Li, S. Chen and S. Zhang, "Integrated Multistage Self-Healing in Smart Distribution Grids Using Decentralized Multiagent," *IEEE Access*, vol. 9, pp. 159081-159092, 2021.

[21] Z. Chu, T.A. Le, H.X. Nguyen, A. Nallanathan and M. Karamanoglu, "A Stackelberg-Game Approach for Disaster-Recovery Communications Utilizing Cooperative D2D," *IEEE Access*, vol. 6, pp. 10733-10742, 2017.

[22] M.H. Rehmani, A. Davy, B. Jennings and C. Assi, "Software Defined Networks-Based Smart Grid Communication: A Comprehensive Survey," *IEEE Commun. Surv. Tut.*, vol. 21, no. 3, pp. 2637-2670, 2019.

[23] S. Shao, W. Gong, H. Yang, S. Guo, L. Chen and A. Xiong, "Data Trusted Sharing Delivery: A Blockchain-Assisted Software-Defined Content Delivery Network," *IEEE Internet Things J.*, vol. 10, no. 14, pp. 11949-11959, 2023.

[24] H. Lin, et al., "Self-healing attack-resilient PMU network for power system operation," *IEEE T. Smart Grid*, vol. 9, no. 3, pp. 1551-1565, 2018.

[25] T. Krause, R. Ernst, B. Klaer, I. Hacker and M. Henze, "Cybersecurity in power grids: Challenges and opportunities," *Sensors-Basel*, vol. 21, no. 18, 2021.

[26] L. Liao, V.C.M. Leung and M. Chen, "An Efficient and Accurate Link Latency Monitoring Method for Low-Latency Software-Defined Networks," *IEEE T. Instrum. Meas.*, vol. 68, no. 2, pp. 377-391, 2019.

[27] M. Wang, Z. Fan, J. Zhou and S. Shi, "Research on Urban Load Rapid Recovery Strategy Based on Improved Weighted Power Flow Entropy," *IEEE Access*, vol. 9, pp. 10634-10644, 2021.

[28] M.E. Baran and F.F. Wu, "Network reconfiguration in distribution systems for loss reduction and load balancing," *IEEE T. Power Deliver.*, vol. v, no. n, pp. 1401-1407, 1992.

[29] "IEC/IEEE International Standard - Communication networks and systems for power utility automation - Part 9-3: Precision time protocol profile for power utility automation,", pp. 1-18, 2016.

[30] S. Wijethilaka and M. Liyanage, "Survey on Network Slicing for Internet of Things Realization in 5G Networks," *IEEE Commun. Surv. Tut.*, vol. 23, no. 2, pp. 957-994, 2021.